\documentclass[preprint]{aastex}

\pdfoutput=1
\usepackage{bm}
\usepackage{ulem}
\usepackage{amsmath}
\usepackage{lineno}

\shorttitle{Kinetic Simulation of slow-mode wave} \shortauthors{Ruan et al.}

\begin{document}

\title{Kinetic Simulation of Slow Magnetosonic Waves and Quasi-periodic Upflows in the Solar Corona}

\author{ Wenzhi Ruan\altaffilmark{1}, Jiansen He\altaffilmark{1,2,*}, Lei Zhang\altaffilmark{2}, Christian Vocks\altaffilmark{3}, Eckart Marsch\altaffilmark{4}, Chuanyi Tu\altaffilmark{1}, Hardi Peter\altaffilmark{5}, Linghua Wang\altaffilmark{1} }

\altaffiltext{1}{School of Earth and Space Sciences, Peking University, Beijing,
100871, China}
\altaffiltext{*}{To whom correspondence should be addressed: jshept@gmail.com}
\altaffiltext{2}{State Key Laboratory of Space Weather, Chinese Academy of Sciences, Beijing 100190, China}
\altaffiltext{3}{Leibniz-Institut f\"{u}r Astrophysik Potsdam, 14482, Potsdam, Germany}
\altaffiltext{4}{Institute for Experimental and Applied Physics, Christian-Albrechts-Universit\"{a}t zu Kiel, 24118 Kiel, Germany}
\altaffiltext{5}{Max Plank Institut f\"{u}r Sonnensystemforschung,Justus-von-Liebig-Weg 3, 37077 G\"{o}ttingen, Germany}


\begin{abstract}

Quasi-periodic disturbances of emission-line parameters are frequently observed in the corona. These disturbances propagate upward along the magnetic field with speeds $\sim100~\rm{km~s}^{-1}$. This phenomenon has been interpreted as evidence of the propagation of slow magnetosonic waves or argued to be signature of the intermittent outflows superposed on the background plasmas. Here we aim to present a new ``wave + flow'' model to interpret these observations. In our scenario, the oscillatory motion is a slow mode wave, and the flow is associated with a beam created by the wave-particle interaction owing to Landau resonance. With the help of a Vlasov model, we simulate the propagation of the slow mode wave and the generation of the beam flow. We find that weak periodic beam flows can be generated owing to Landau resonance in the solar corona, and the phase with strongest blueward asymmetry is ahead of that with strongest blueshift by about 1/4 period. We also find that the slow wave damps to the level of 1/e after the transit time of two wave periods, owing to Landau damping and Coulomb collisions in our simulation. This damping time scale is similar to that resulting from thermal-conduction in the magnetohydrodynamics regime. The beam flow is weakened/attenuated with increasing wave period and decreasing wave amplitude since Coulomb collision becomes more and more dominant over the wave action. We suggest that this ``wave + flow" kinetic model provides an alternative explanation for the observed quasi-periodic propagating perturbations in various parameters in the solar corona.

\end{abstract}

\keywords{solar corona --- waves --- disturbances}


\section{Introduction}

Quasi-periodic upward propagating intensity disturbances are frequently observed along the magnetic field structure in the solar corona.
The disturbances were observed in polar coronal plumes \citep{Ofman1997ApJL, DeForest1998ApJL, Banerjee2000SoPh}, and in the coronal fan loops at the edges of active regions \citep{Berghmans1999SP, DeMoortel2000A&A}.
The parameters of the propagating intensity disturbance (PIDs) are summarised by \citet{DeMoortel2009SSR} as follows: oscilation period of $ 145-550 ~ \rm s $, propagation speed of $ 45-205 ~ \rm km \ s^{-1} $, relative intensity amplitude of $ 0.7\% - 14.6\% $, detected wave lengths of $ 2.9-23.2 ~ \rm Mm $.
These propagating intensity disturbances (PIDs) are often accompanied with Doppler-shifts in the spectral lines.
The phenomenon of PIDs had been almost universally interpreted as the propagation of slow magnetosonic waves \citep{Nakariakov2000A&A}.
The slow-mode waves may play an important role in the heating of the chromosphere, the generation of solar spicules, and the development of coronal loops \citep{Hollweg1982SoPh, Shibata1982SoPh, Poter1994APJ, Kumar2006A&A, Yang2013ApJ, Liu2014RAA}. The slow-mode waves in the corona may also be connected with (driven by) the spicule / jet flows in the chromosphere \citep{Jiao2015ApJ, Samanta2015ApJ}. Another important piece of evidence supporting the slow-mode wave scenario comes from the good correlation between intensity and Doppler-shift variations as derived from the spectroscopic observations by the \textit{Extreme-ultraviolet Imaging Spectrometer} (EIS) onboard \textit{Hinode} \citep{Wang2009A&A}.

However, this interpretation is challenged by some authors \citep{DePontieu2010ApJ, Tian2011ApJL} , who claimed that the quasi-periodic ``Red Minus Blue" (R-B) asymmetries found in the spectral lines of intensity disturbance region are signatures of quasi-periodic upflows.
Hence, the debate on whether, the slow wave or the intermittent outflow, corresponds to the real nature of the disturbances is initiated and continues.
The slow wave and the flow are thought to be related to different dynamic processes.
The intermittent outflows inferred at the edges of the active regions are thought to be the possible source of the solar wind \citep{Sakao2007Sci, Harra2008ApJ, Hara2008ApJ,  McIntosh2009ApJ, He2010A&A, Warren2011ApJ}.
Ascertaining the nature of the intensity disturbances is crucial for clarifying the physical processes in the dynamic solar atmosphere.

The excess line width enhancement beyond the pure-wave model may be explained by the superposition of waves on the background uncoupled plasmas, although the resultant frequency of the line width oscillation may be twice original one \citep{Verwichte2010ApJ, Wang2012ASPC}.
A dual magnetohydrodynamic (MHD) scenario/model has also been proposed to launch the slow mode waves as excited by the quasi-periodic flows at the footpoints of the magnetic flux tubes \citep{Nishizuka2011ApJ, Ofman2012ApJ, Wang2013ApJ}.
In this dual model, slow mode waves are gradually decoupled from the upflows and propagate at lager speed to higher altitude than the latter one.
The PIDs in the coronal strands with extended length seem to be the signature of slow mode waves rather than quasi-periodic flows in their model.

It seems that these two interpretations are incompatible in the MHD regime.
Nevertheless, it is known that Landau damping of the slow wave can generate a beam and associated plasma flow.
Consequently, the wave and flow may contemporaneously exist self-consistently, and the two diverse interpretations of the observed intensity disturbance can be compatible.
To reconcile these two interpretations, we present a new senario and call it ``wave + flow" kinetic scenario, which involves both the slow wave and the beam created by Landau resonance with the waves.
We suggest that the observed R-B asymmetries in the spectral line may be the signal of beam related flows.
We use kinetic simulation to test this scenario, and we reproduce the weak beam component in the ion velocity distributions within the context of slow wave propagation.
The influence of Coulomb collisions is also taken into account in this kinetic simulation model.

The kinetic simulation model is described in Sect.~2.
In Sect.~3, the simulation setup and results are presented, including the introduction of slow-mode oscillation at the bottom of simulation region and the corresponding response of the proton velocity distribution function (VDF) in terms of different moments (density, bulk velocity, and R-B asymmetry).
The variations of the parameters of the slow mode wave and their kinetic effects on the R-B asymmetry of the proton VDF are investigated.
The damping mechanism of the slow mode wave in our simulation is discussed as well at the end of this section.
We conclude our paper with a discussion of the applicability of our ``wave + flow" model in Sect.~4.
The limitation of our work is also discussed in this section.


\section{Simulation model}

In this section, we briefly describe the model we used in our kinetic simulation.
This model was first introduced by \citet{Vocks2001GRL}.
This model is based on the Vlasov equation,
\begin{equation}
\frac{\partial f}{\partial t} + (\bm{v \cdot \nabla})f +
[\bm{g} + \frac{q}{m}(\bm{E} + \bm{v \times B})] \bm{\cdot \nabla}_v f =
\Big( \frac{\delta f}{\delta t} \Big)_{\rm{Coul}},
\end{equation}
where $ \bm{B} $ is the magnetic field, $ \bm{E} $ is the electric field, $ q $ is the particle's charge and $ m $ its mass, and   $ \bm{g} $ is the gravitational acceleration.
The term on the right-hand side denotes the Coulomb collisions.
The velocity distribution function (VDF) $ f(\bm{r}, \bm{v}, t) $ depends on three velocity and three spatial coordinates, and on time.
The computational cost is high to solve the Vlasov equation in the complete six dimensional phase space.
To simplify, it is necessary to reduce the dimensions of $ f $.
Since the ion gyroperiod is short compared to other characteristic timescales, it is resonable to assume gyrotropy \citep{Vocks2001GRL}.
Therefore, the number of velocity coordinates can be reduced from three to two: $ \bm{v} \to (v_{\|}, v_{\bot}) $ \citep{Vocks2002ApJ}.
In this kinetic model, the ion velocity component parallel to the background magnetic field is of primary interest, and only waves propagating in the parallel direction are considered.
Accordingly, a ``reduced VDF" \citep{Marsch1998NPG, Vocks2001GRL, Vocks2002ApJ, Vocks&Marsch2002ApJ} is introduced in this model and obtained by integrating over the velocity component perpendicular to the background magnetic field:
\begin{equation}
F_k(v_{\|}) =
2 \pi \int_{0}^{\infty} v_{\bot}^{2k+1} f(v_{\|}, v_{\bot}) d v_{\bot}
\qquad
k = 0, 1, 2, \ldots \quad.
\end{equation}

A Vlasov equation for the reduced VDFs can be derived by intergrating over $ v_{\bot} $ in the same way as in the definition of $ F_k $ (Equation (2)):
\begin{eqnarray}
\frac{\partial F_k}{\partial t} + v_{\|} \frac{\partial F_k}{\partial s} +
( \frac{q}{m} E_{\|} - g \cos \psi ) \frac{\partial F_k}{\partial v_{\|}} +
\nonumber \\
\frac{1}{2A} \frac{\partial A}{\partial s}
[\frac{\partial F_{k+1}}{\partial v_{\|}} + 2v_{\|} (k+1) F_k] =
\Big( \frac{\delta F_k}{\delta t} \Big)_{\rm{Coul}},
\end{eqnarray}
where $ E_{\|}(s) $ is the electric field component parallel to the magnetic field, $ A(s) $ is the cross sectional area of the magnetic flux tube, and $ \psi(s) $ is the angle between the magnetic field and the direction normal to the solar surface.
The term on the right-hand side which denotes the Coulomb collisions, is derived from the Fokker-Planck collision term, for which the formulation of \citet{Ljepojevic1990RSPSA} is used.
Only the spatial coordinate $ s $ parallel to the magnetic field is considered in this model.
The Vlasov equation for the reduced VDFs $ F_k $ depends on $ F_{k+1} $, and therefore a cut off is needed.
The assumption,
\begin{equation}
F_k (v_{\|}) = k! (2 v_{th, \bot}^2)^{k-1} F_1 (v_{\|}),
\end{equation}
is applied in this model.
This assumption is exact for a Maxwellian in $ v_{\bot} $, and it is easily to satisfied in the collisional plasma.

Equation (3) for $ k = 0, 1 $ is used to determine the governing equations in this model.
The particle density $ N $, drift velocity $ v_d $, parallel temperature $ T_{\|} $, and heat flux $ q_{\|} $ can be obtained from $ F_0 $.
The perpendicular temperature $ T_{\bot} $ and the heat flux $ q_{\bot} $ can be obtained from $ F_1 $.
The Vlasov equations for the reduced VDFs are only used to describe the ion kinetics.
Electrons are dealt with in the fluid approximation.

$ E_{\|} $ in this model is determined from the electron momentum equation
\begin{equation}
m_e \frac{d \bm{v}_e}{d t} =
- \frac{1}{N_e} \nabla p_e -
e (\bm{E} + \bm{v}_e  \bm{\times} \bm{B}).
\end{equation}
Considering that the ions and electrons nearly have the same bulk drift velocity, the term on the left-hand side can be neglected because the electron mass is quite small.
Under the approximation of the quasi neutrality condition $ e N_e = \sum_j q_j N_j $, we can deduce the form of $ E_{\|} $:
\begin{equation}
E_{\|} = - \frac{1}{e \sum_j q_j N_j}
\frac{\partial}{\partial s} \Bigg( k_B T_e \sum_j q_j N_j \Bigg).
\end{equation}


\section{Simulation setup and results}


\subsection{Introducing the slow-mode waves}

The simulation results on the evolution and kinetic effect of the slow wave propagating in a magnetic flux tube are presented in this section.
We assume that the lower boundary of the magnetic flux tube is located in the low corona, and the flux tube is normal to the surface of the sun and has a constant cross-sectional area.
The model plasma in the flux tube consist of protons and electrons.
The gravity is considered in our simulation, with the gravitational acceleration     $ g = -274 ~ \rm km ~ s^{-2} $, and thus a gradient exists in the profile of the proton's number density.
At the beginning of the simulation, the number density of proton $N_0$ is set so at the bottom boundary $ N_0 (h = 0 ~ \rm Mm) = 1 \times 10^8 ~ \rm cm^{-3} $ and at the top boundary $ N_0 (h = 100 ~ \rm Mm) = 1.9 \times 10^7 ~ \rm cm^{-3} $.
The initial temperature of the plasma in the whole magnetic flux tube is $ T_0 = 1 ~ \rm MK $.
The plasma is initially in an equilibrium state.

We introduce the slow wave into the flux tube by setting the parameters of the plasma at the lower boundary.
Here we set:
\begin{equation}
N_b(t) = [1 + \epsilon \cos(2 \pi t / \tau)] N_b(0),
\end{equation}
\begin{equation}
v_b(t) = \epsilon \cos (2 \pi t / \tau) c_s,
\end{equation}
\begin{equation}
T_b(t) = [1 + (\gamma - 1) \cos(2 \pi t / \tau)] T_b(0),
\end{equation}
where $ N_b(t) $, $ v_b(t) $ and $ T_b(t) $ are the number density, drift velocity and temperature respectively at the bottom boundary at simulation time $ t $.
The plasma at this boundary is assumed to be in a state of thermal equilibrium, the velocity distribution of which is Maxwellian.
The constant $ \epsilon $ is the relative amplitude of the slow wave, $ \tau $ is its period and $ \gamma = 5/3 $ is the adiabatic exponent.
As an example, the amplitude is chosen as $ \epsilon = 0.1 $, and the period is chosen as      $ \tau = 120 ~ \rm s $.
The acoustic speed is set to $ c_s = 166 ~ \rm km \ s^{-1} $, which is calculated theoretically from isotropic MHD assuming the processes are adiabatic.
In this paper, a positive velocity means that it is in the upward direction.

\begin{figure}
\centering
\includegraphics[width=12cm]{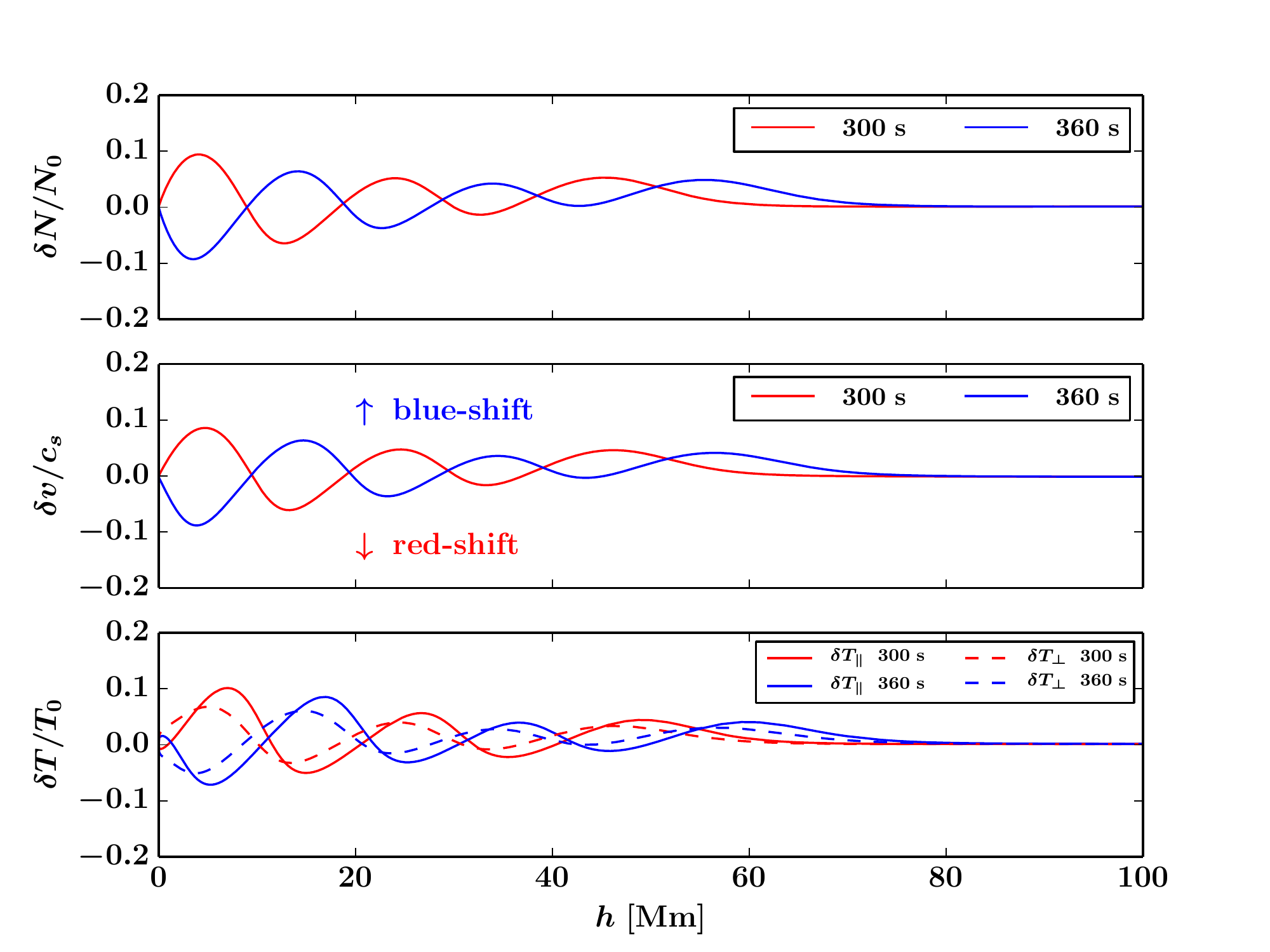}
\caption{Height profiles of density (top panel), drift velocity (middle panel), parallel temperature (bottom panel, solid line) and perpendicular temperature (bottom panel, dashed line) at $ t = 300 ~ \rm s $ (red line) and $ t = 360 ~ \rm s $ (blue line). }
\label{fig:digit}
\end{figure}

Fig.~1 displays the height profiles of the number density, drift velocity, parallel temperature and perpendicular temperature at $ t = 300 ~ \rm s $ and $ t = 360 ~ \rm s $.
In Fig.~1, $ \delta N $, $ \delta v $ and $ \delta T $ denote the fluctuations away from the initial quantity, i.e. $ \delta N(h) = N_p(h) - N_0(h) $, where $ N_p(h) $ is the number density at $ t = 300 ~ \rm s $ or $ t = 360 ~ \rm s $ and $ N_0(h) $ is the number density at $ t = 0 ~ \rm s $.
$ T_0(h) $ is the height profile of the temperature at $ t = 0 ~ \rm s $.

In Fig.~1, periodic variations can be found in the profiles of number density, drift velocity and temperature.
The third peak of the drift velocity arrives the height $ h = 4.7 ~ \rm Mm $ at $ t = 300 ~ \rm s $ and the height $ h = 14.7 ~ \rm Mm $ at $ t = 360 ~ \rm s $.
The propagation speed of velocity perturbation is thus about $ 167 \ \rm km \ s^{-1} $, which is approximate to the acoustic speed of plasma $ c_s $ as calculated theoretically under the MHD regime.
Accordingly, we consider that the slow wave is successfully launched introduced into the magnetic flux tube.

The temperature of plasma in Fig.~1 is anisotropic, the amplitude of          $ \delta T_{\parallel} $ is larger than the amplitude of $ \delta T_{\perp} $, and the phase of $ \delta T_{\parallel} $ is ahead of the phase of                $ \delta T_{\perp} $.
$ T_{\parallel} $ and $ T_{\perp} $ are coupled by the Coulomb collisions here.
As a result, a perturbation can be found in the profile of $ T_{\perp} $.
However, the Coulomb friction is not so strong to establish an isotropic temperature distribution, so the plasma remains anisotropic in temperature.


\subsection{The kinetic effect of slow-mode wave}

\begin{figure}
  \centering
  \includegraphics[width=12cm]{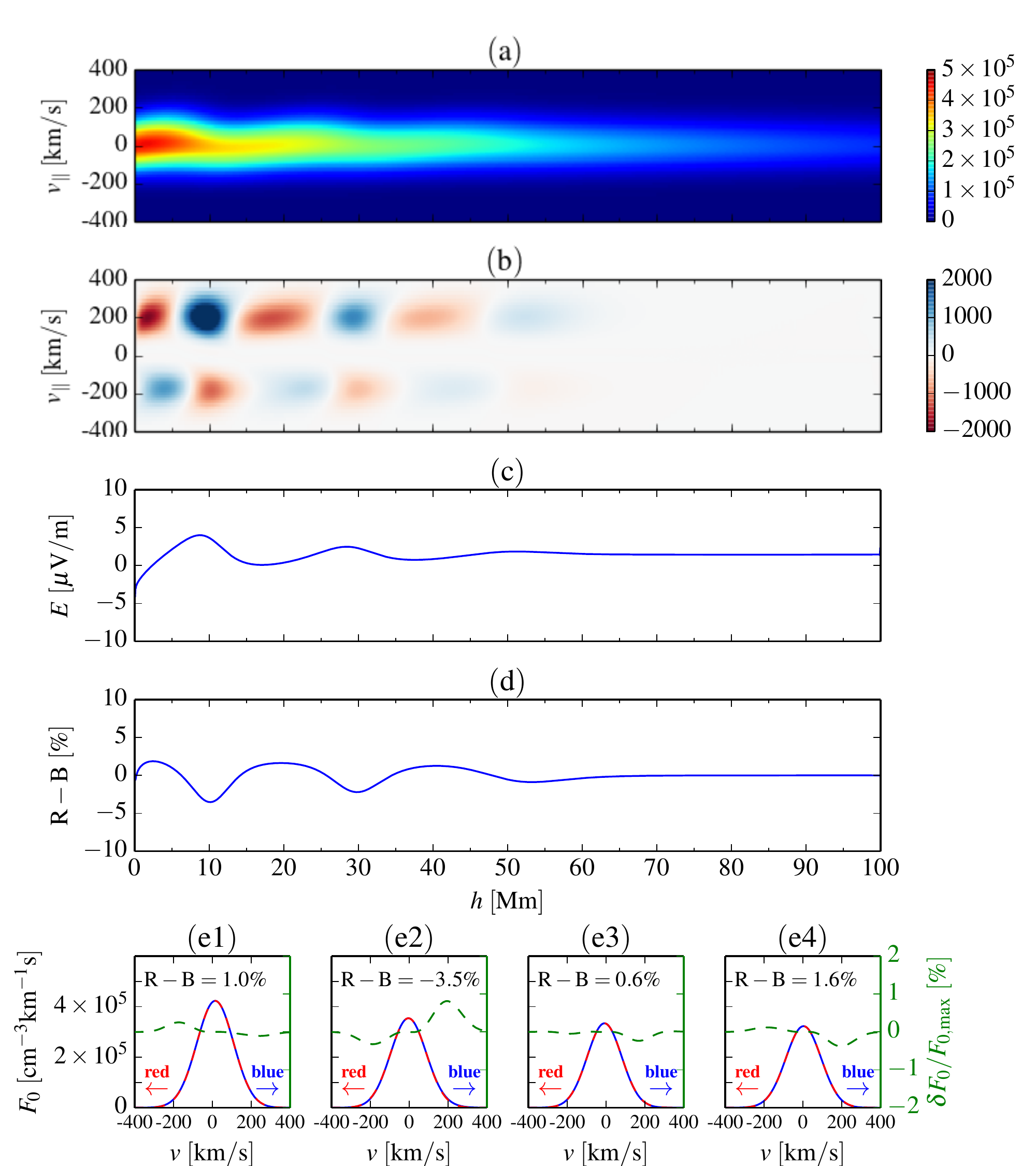}
  \caption{Plasma parameters at $ t = 300 ~\rm s $. (a) Parallel velocity distribution fuction $ F_0(h,v) $ of the protons. (b) The distribution of $ \delta F_{0,\textrm{non-Mxw}} (h,v) = F_0(h,v) - F_{0, \textrm{fit}}(h,v) $, where $ F_{0, \textrm{fit}}(h,v) $ is the Maxwellian fitting of $ F_0(h,v) $. (c) The electric field $ E $. (d) The values of R-B. (e1)$ \sim $(e4) Velocity distributions $ F_0(v) $ (blue solid lines), the Maxwellian fitting of the distributions (red dashed lines), and the correspongding values of $\delta F_{0,\textrm{non-Mxw}}(v) / F_{0, \textrm{max}} $ (green dashed lines) at $ h = 5, 10, 15,20 ~ \rm Mm $, where $ F_{0, \textrm{max}} $ is the maximum value $ F_0(v) $. The R-B values of the distributions are listed in the plots.}
  \label{fig:digit}
\end{figure}

The proton kinetics, i.e. characteristics of proton velocity distribution, in association with the slow wave is investigated in this sub-section.
The proton VDF $ F_0(h,v) $ along the magnetic flux tube is displayed in Fig.~2a.
The perturbation is more obvious in the half part of phase space where the parallel velocity $ v_{\parallel} > 0 $ and less obvious in the other half part where $ v_{\parallel} < 0 $.
This asymmetric pattern in $ F_0(h,v) $ is determined by the nature of slow wave.
The perturbations of drift velocity, number density and temperature are nearly in the same phase for the slow-mode wave.
When the bulk drift velocity becomes larger, the number density become larger and the temperature become higher.
As a result, the velocity distribution is wider when the bulk drift velocity $ v_{d} > 0 $ and narrower when $ v_{d} < 0 $.
So the perturbation is more obvious in the velocity domain with $ v_{\parallel} > 0 $.

To evaluate the deviation of VDF from the Maxwell distribution and reveal the beam component flow, a function $ \delta F_{0,\textrm{non-Mxw}}(h,v) $ is defined as a measure of the non-thermal state:
\begin{equation}
 \delta F_{0,\textrm{non-Mxw}}(h,v) = F_{0}(h,v) - F_{0, \textrm{fit}}(h,v),
\end{equation}
where $ F_{0, \textrm{fit}}(h,v) $ is the Maxwellian fitting of $ F_0(h,v) $.
The values of $ \delta F_{0,\textrm{non-Mxw}} (h,v) $ at $ t = 300 ~ \rm s $ are displayed in Fig.~2b.
The VDF $ F_0(h,v) $ of protons deviates from the Maxwell distribution periodically.
In the regions $ -300 ~ \rm km \ s^{-1} $ $ < v_{\parallel} < $ $ -100 ~ \rm km \ s^{-1} $ and $ 100 ~ \rm km \ s^{-1} $ $ < v_{\parallel} < $ $ 300 ~ \rm km \ s^{-1} $, the deviation from a Maxwell distribution is more obvious, for the reason that the frequency of collisions between local protons there and the major population of protons with small $ |v_{\parallel}| $ is lower, while the number density of proton is not very small there.
Stronger beam component flows can be found in the region $ 100 ~ \rm km \ s^{-1} $ $ < v_{\parallel} < $ $ 300 ~ \rm km \ s^{-1} $, since the particles in this velocity range move nearly in phase with the propagating wave electric field $ E $.
These particles can involves in Landau resonance with slow-mode wave.

Fig.~2e illustrates the VDFs at four different heights with four different phases: (e1), $ h = 5 ~ \rm Mm $, $ v_d $ is at the maximum; (e2), $ h = 10 ~ \rm Mm $,    $ v_d = 0 $, $ \partial v_d / \partial h < 0 $; (e3), $ h = 15 ~ \rm Mm $, $ v_d $ is at the minimum; (e4), $ h = 20 ~ \rm Mm $, $ v_d = 0 $, $ \partial v_d / \partial h > 0 $.
The Coulomb friction can not be neglected, thus the VDFs (blue solid lines) in Fig.~2b are not far from the Maxwellian ion distributions (red dashed lines).
Nevertheless, we can find a weak asymmertry in these VDFs whenever we perform a R-B profile asymmetry analysis on them.
The non-thermal components as deviating from the Maxwellian distribution are plotted with green dashed lines in Fig.~2e.
We perform the R-B analysis in the similar way as \citet{Verwichte2010ApJ} did.
We treat the part with $ v < v_d $ as red wing and the part with $ v > v_d $ as blue wing in our work.
The results of the R-B analysis at $ 120 - 300 ~ \rm km \ s^{-1} $ off the velocity center ($ v_{d} $) are shown in Fig.~2b.
The R-B values above/below 0 indicate redward/blueward asymmetries.
Among these four phases of slow wave, the R-B estimate with maximum absolute value is $ -3.5 \% $ (panel e2 of Fig.~2).

Hight profiles of the electric field and the R-B values are shown in Fig.~2c and Fig.~2d respectively.
The peaks of blueward asymmetry (the minimum of R-B) profile are located  near the peaks of electric field $ E $.
At the heights where the blueward asymmetry is peaked, a balance is achieved between the formation of beam flows caused by Landau resonance and the destruction of beam flows dominated by Coulomb friction.
The traditional particle trapping scenario in Landau resonance is not applicable here, as the mean free path of proton is far less than the wave length.

\begin{figure}
  \centering
  \includegraphics[width=12cm]{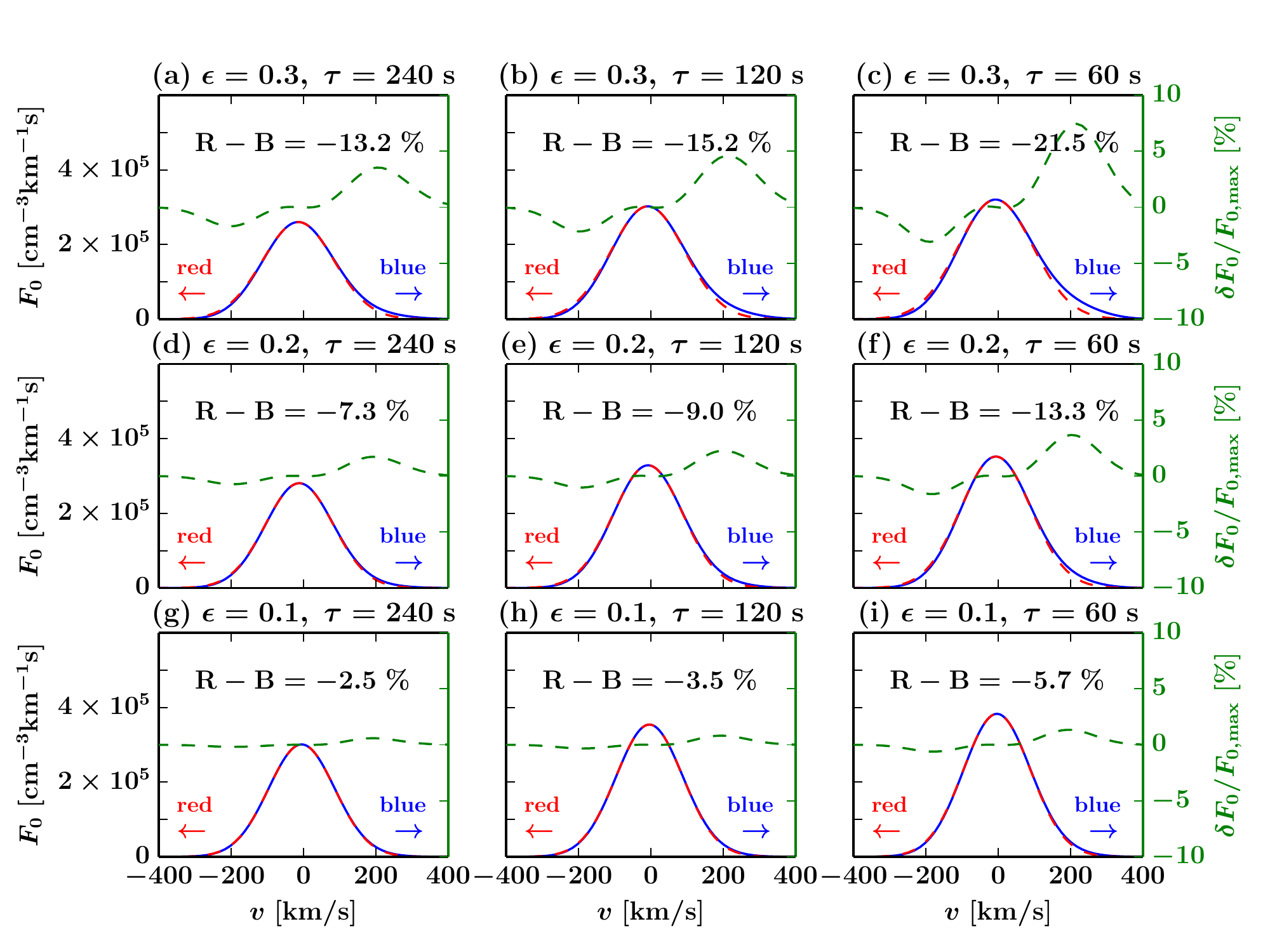}
  \caption{VDFs of simulations at the phase of $E_\parallel\rm{=}E_{\rm{max}}$ with different amplitudes ($\epsilon\rm{=}0.1, 0.2, 0.3$) and periods ($\tau\rm{=}\rm 60 ~ s, 120 ~ s, 240 ~ s$). The blue solid lines denote the $F_0(v)$, the red dashed lines denote the Maxwellian fitting of $F_0(v)$, and the green dashed lines denote the values of $\delta F_{0,\textrm{non-Mxw}}(v) / F_{0, \textrm{max}} $, where $  F_{0, \textrm{max}} $ is the maximum value of the $F_0(v)$. Here the height and time corresponding to the phase of $E_\parallel\rm{=}E_{\rm{max}}$ are selected as $h\rm{=}\rm 5 ~ Mm, 10 ~ Mm, 20 ~ Mm $ and $ t\rm{=}\rm 270 ~ s, 300 ~ s, 360 ~ s $ for different periods $ \tau\rm{=}\rm 60 ~ s, 120 ~ s, 240 ~ s $, respectively.}
  \label{fig:digit}
\end{figure}

To evaluate the influence of wave parameters on the periodic beam flow formation, nine simulations are run with different amplitude and period of the slow wave.
The wave amplitudes $ \epsilon = 0.1, 0.2, 0.3 $ and the wave periods           $ \tau = \rm 60 ~ s, 120 ~ s, 240 ~ s $ are chosen.
The VDFs of these nine simulations at the phase with the most blueward asymetry and corresponding R-B value, and the values of $\delta F_{0,\textrm{non-Mxw}}(v) / F_{0, \textrm{max}} $ are illustrated in Fig.~3, where $F_{0, \textrm{max}}$ is the peak value of $F_0$.
The VDF at the phase of obivous beam flows in each simulation is plotted.
The amplitudes and periods are listed on the top of each plot.
The damping rate of the slow wave is determined by the wave period.
We note that the waves with longer periods can propagate to higher positions, and the Landau resonance can happen in a larger spatial range.
As a result, the most obvious beams tend to appear at higher positions for the waves with longer wave periods.
For a comparison between simulation results with different wave periods ($ \tau\rm{=} \rm 60 ~ s, 120 ~ s, 240 ~ s $), VDFs at the similar phase with maximum $E_\parallel$ (e.g., $E_\parallel\rm{=}E_{\rm{max}}$ at $ h\rm{=}\rm 5 ~ Mm, 10 ~ Mm, 20 ~ Mm $ and $ t\rm{=}\rm 270 ~ s, 300 ~ s, 360 ~ s $ for $ \tau\rm{=}\rm 60 ~ s, 120 ~ s, 240 ~ s $, respectively) are selected and shown in Fig.~3.
Obvious beam flows tend to appear in the simulations with decreasing wave periods from left to right in Fig.~3 and increasing wave amplitudes from bottom to top in Fig.~3.
As the wave period shortens or the wave amplifies, the plasma will undergo more intensive charge separation, and therefore the electric field will strengthen.

\begin{figure}
  \centering
  \includegraphics[width=12cm]{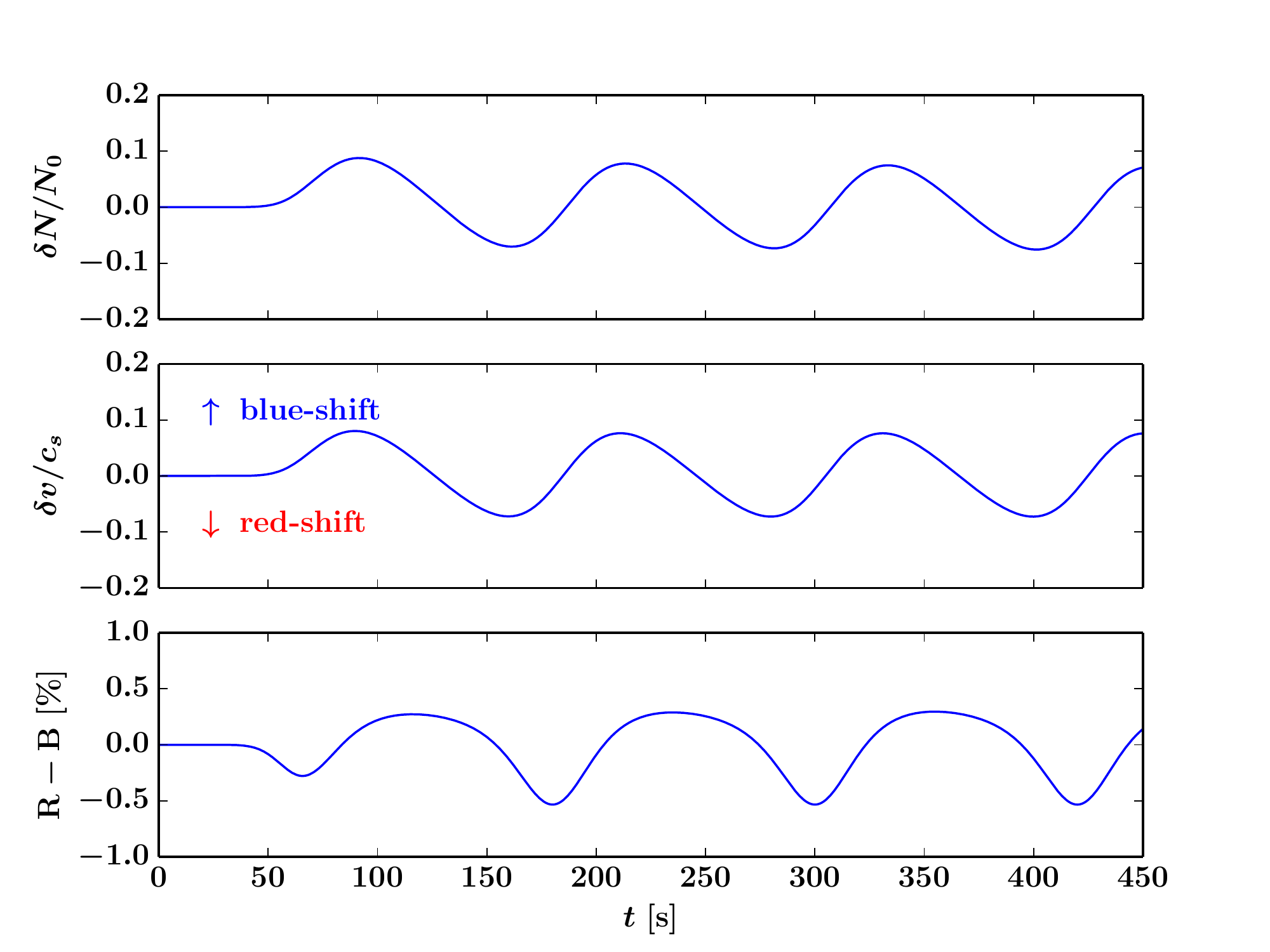}
  \caption{Time evolution of the number density, the drift velocity and the R-B values at $ h = 10 ~ \rm Mm $. The positive R-B values indicate redward asymmetries.}
  \label{fig:digit}
\end{figure}

Time evolution of the number density, the drift velocity and the R-B asymmetry at $ h = 10 ~ \rm Mm $ are displayed in Fig.~4 for comparison with the observed results.
The strongest blueward asymmetries appear ahead of the phase with strongest blue shift by about 30s (1/4 of the wave period).
It is different from the classical intermittent outflow scenario \citep{DePontieu2010ApJ, Tian2011ApJL}, in which the strongest blueward asymmetries appear at the time when the strongest blueshift appear.
This difference arises because these authors assumed the blueward asymmetry is caused by the superposition of intermittent fast flow on the background plasmas nearby but unresolved under current spatial resolution from observations.
In their scenario, the weak correlation between Doppler-shift and R-B asymmetry from observations is speculated to be caused by strong noise of R-B asymmetry.
We suggest an alternative explanation for the weak correlation: phase difference between Doppler shift and R-B asymmetry as a result of slow wave kinetic effects.


\subsection{Damping of slow-mode wave}

Here we discuss the damping mechanism.
As shown in Fig.~1, the slow waves damp after the propagation of several wavelengths.
We suggest that the damping of slow wave consists of the concurrent twofold processes: (1) formation of heat flux contributed from non-thermal velocity tail owing to Landau resonance with slow wave; (2) thermalization of non-thermal tail particles (dissipation of heat flux) by Coulomb collisions.
This mechanism behaves similarly to the thermal conduction in MHD or fluid equations.
This similarity of damping evolution can be demonstrated by comparing the height profile of wave amplitude in the simulation with the theoretical height profile of amplitude as derived from the fluid equations.

To simplify this comparision, gravity and gravitational stratification of plasma are not considered in the test kinetic simulation and the fluid equations.
A new test simulation without gravitational stratification was run.
The background number density of proton is $ N_0 = 1 \times 10^8 ~ \rm cm^{-3} $, and the background temperature is $ T_0 = 1 ~ \rm MK $ in the simulation and the fluid equations.
For the energy equation, the form in \citet{Owen2009A&A} was used.
The fluid equations read:
\begin{equation}
\frac{\partial \rho_p}{\partial t} + \frac{\partial \rho_p v_p}{\partial h} = 0,
\end{equation}
\begin{equation}
\frac{\partial \rho_p v_p}{\partial t} + v_p \frac{\partial \rho_p v_p}{\partial h} =
- \frac{\partial P_p}{\partial h} + \rho_{qp} E,
\end{equation}
\begin{equation}
\frac{\partial \epsilon_p}{\partial t} + v_p \frac{\partial \epsilon_p}{\partial h} =
- (\gamma - 1) \epsilon_p \frac{\partial v_p}{\partial h} +
\frac{1}{\rho_p} \frac{\partial}{\partial h}(\kappa_p \frac{\partial T_p}{\partial h}),
\end{equation}
\begin{equation}
\epsilon_p = \frac{P_p}{(\gamma - 1) \rho_p},
\end{equation}
where $ \rho_p $ is the proton density, $ v_p $ the proton drift velocity, $ P_p $ the proton pressure, $ \rho_{qp} $ the proton charge density, $ \epsilon_p $ the specific internal energy of proton per unit mass, $ \gamma = 5/3 $ the ratio of the specific heats, and $ T_p $ the proton temperature.
The electric field $ E $ is the same as the from of $ E_{\parallel} $ in Equation (6).
The coefficient of thermal conduction $ \kappa_p = 434 \rm ~ W ~ m^{-1} ~ K^{-1} $ is estimated from the ratio $ {Q_p(h)}/{(\partial T_p(h) / \partial h)} $ in our simulation, where $ Q_p(h) $ is the heat flux of the protons.
The Coulomb friction between protons and electrons is not included in Equations (11) to (13), as electrons are not dealt with as kinetic particles in the simulation model.

We linearize the Equations (11) - (13), and calculate the theoretical height profile of the velocity amplitude using the same method as \citet{Owen2009A&A}.
Theoretical velocity amplitude is $ \delta v(h) / c_s = (\delta v(0) / c_s)  \textrm{exp} (- k_i h ) $, where $ k_i = 2.88 \times 10^{-8} ~ \rm m^{-1}$.
The theoretical drift velocity amplitude given by Equations (11) - (13) (black dashed line) and the drift velocity of simulation at $ t = 390 ~ \rm s $ (blue solid line) are shown in Fig.~5a.
For the drift velocity of simulation, a periodic average drift velocity is substracted.
The simulation result agrees well with the predicted one, which means that the damping of slow wave is dominated by the thermal conduction if the gravity and gravitational stratification of plasma are not considered.

The drift velocities of the first simulation (with gravitational stratification) and the second simulation (without gravitational stratification) at $ t = 390 ~ \rm s $ are shown in Fig.~5b.

\begin{figure}
  \centering
  \includegraphics[width=12cm]{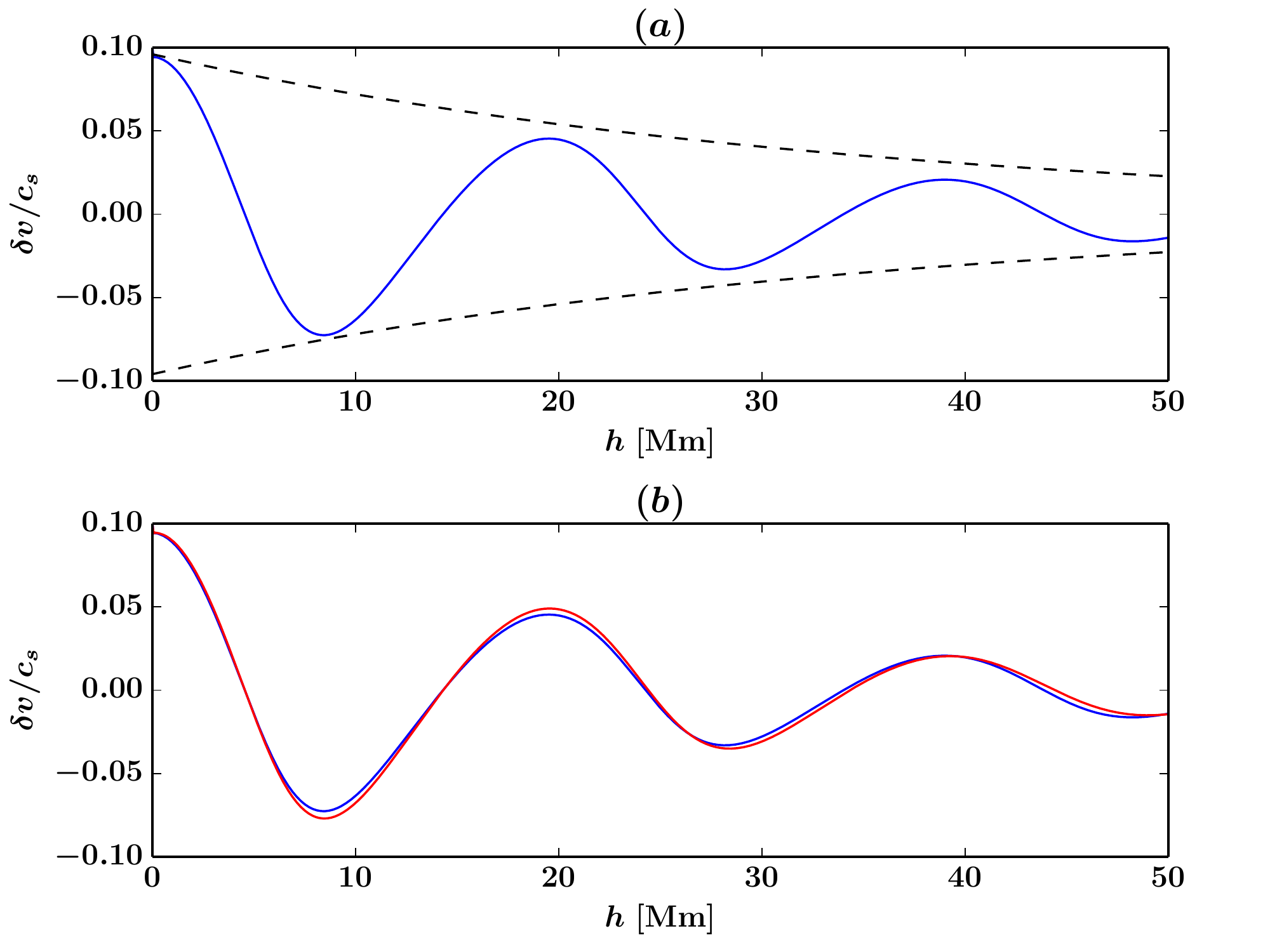}
  \caption{(a) Theoretical drift velocity amplitude (black dashed line) from fluid theory and the drift velocity (blue solid line) at $ t = 390 ~ \rm s $ of simulation without gravitational stratification. (b) Drift velocity of simulation without gravitational stratification (blue solid line) and drift velocity of simulation with gravitational stratification (red solid line).}
  \label{fig:digit}
\end{figure}

When considering the gravitational stratification, the wave amplitude is found to not increase rapidly.
The scenario differs from \citet{Owen2009A&A}'s results, where the stratification leads to growth of the amplitude of velocity exponentially as Eq.~(15).
\begin{equation}
v = \hat{v} \ \textrm{exp}  \frac{1}{2  H},
\end{equation}
where $ H $ is the integrated gravitational scale height.
However, from Equation (13) we know that, the influence of thermal conduction grows rapidly when the proton density $ \rho_p $ decreases, as $ \kappa_p $ does not vary too much.
The growth of damping rate owing to thermal conduction counteracts the influence of gravitation in our simulation.
As a result, the height profile of drift velocity does not vary too much between the options with and without considering the gravitational stratification.

We note that which (the stratification or the thermal conduction) dominates the propagation evolution of slow wave is determined by the parameters such as plasma density and wave period.
For example, in \citet{DeMoortel2004A&A}, the influence of stratification is stronger, as the wave period is longer and the plasma density is larger in their work.
The other thing we need to emphasize is that kinetic effects of electrons are not dealt with in our simulation.
However, if we consider again the kinetic effects of electrons, the slow wave is expected to damp faster, as the thermal conduction of electrons is much larger than that of the protons.


\section{Discussion and conclusion}

To interpret the observed quasi-periodic disturbances in emission intensity, Doppler shift, line width, and R-B asymmetry, we present a new ``wave + flow" scenario.
In our scenario, the oscillation is a slow mode wave, and the flow is due to beam component created by the kinetic effect of Landau resonance.
We suggest that the quasi-periodic R-B asymmetries found in the spectral lines in intensity disturbance regions may be the signatures of flows.
However, we do not know whether they can be generated in the low corona, where the plasma is semi-collisional.
Therefore, we test our scenario by simulating the propagation of the slow wave in a magnetic flux tube with a kinetic model.

In our simulations, the plasma in the flux tube consists of protons and electron fluid.
Weak periodic beam components are found in the velocity distributions of the protons.
The formation of the beam is caused by Landau resonance between protons and slow waves, which is partially counteracted by Coulomb collisions between protons.
The signatures of beam flows and R-B asymmetries are periodic as well.
The R-B asymmetry with strongest blue wing enhancement appears at the height where the electric field and the gradient of the proton number density are at their maxima.
When we envision to sit at a constant height and watch the time variation, the strongest blueward asymmetry (minimum R-B) appears before the time of the strongest blueshift (maximum drift velocity).
This phase relation between R-B asymmetry and drift velocity is different from the classical scenario of intermittent outflow \citep{DePontieu2010ApJ}, in which the strongest blueward asymmetry appears at the time of strongest blueshift.
The phase difference between R-B asymmetry and Doppler shift in our model may be one of the reason for the weak correlation found between them from observations.
The other thing of interest is that the slow waves damp fast.
We suggest that this damping may result from concurrent twofold processes: (1) formation of heat flux (non-thermal particles) by Landau resonance; (2) dissipation of the heat flux (thermalization of non-thermal tail in velocity distribution) by Coulomb collisions.
This idea is corroborated by comparing the damping profiles from our kinetic model with those obtained from fluid equations with thermal conduction.

The conclusions are given as follows:
(1) Weak periodic beam components were generated owing to the Landau resonance in our simulation;
(2) The strongest blueward asymmetry appears before the time of strongest buleshift by 1/4 wave period;
(3) The main damping mechanism of slow wave in our simulation is the kinetic process related to thermal conduction;
(4) Our ``wave + flow" scenario may be a viable explanation for the observed quasi-periodic intensity disturbances in the solar corona.

This paper concentrated on whether a beam can be generated in the normal environment of the low corona.
Actually, the quasi-periodic intensity disturbances were observed in the spectra of heavy ions like Fe XII, Fe XIII etc.
To further test our scenario is to simulate the behaviors of such heavy ions, besides the behaviors of protons.
However, the non-thermal turbulent fluctuations are important when studying the kinetic behaviors of heavy ions, as the non-thermal width of the spectral lines is larger than the thermal width of the heavy ion's velocity distribution in the low corona.
But it is difficult to introduce a turbulent fluctuation in the simulation, since the model we use is one-dimensional in velocity and in physical space, but the turbulent fluctuation is a two-dimensional or three-dimensional phenomenon in physical space.

For heavy ions, the thermal velocity is much smaller than the local sound speed, and thus few ions can resonate with the slow mode wave.
However, after considering the turbulent fluctuations, the profiles of the heavy ion velocity distribution will become wider, and more ions can be in the region of resonance with slow waves.
Whether beam flows can be generated or not in this environment needs to be tested with the help of higher dimensional kinetic simulation.

\begin{acknowledgements}
{\bf Acknowledgements:} This work at Peking University is supported by NSFC under contracts 41222032, 41174148, 41574168, 41231069, 41274172, 41474148, and 41421003. JSH is also supported by National Young Top-Notch Talent Program of China.
\end{acknowledgements}

\bibliographystyle{apj}
\bibliography{references}

\end{document}